\documentclass[prl,reprint,amsmath,amssymb,aps,preprintnumbers, longbibliography]{revtex4-1}

\usepackage[colorlinks=true,linkcolor=black, citecolor=black,
urlcolor=black]{hyperref}

\usepackage{multirow,graphics}
\usepackage{enumerate}
\usepackage{amstext}
\usepackage{amssymb}
\usepackage{amsmath}
\usepackage{graphicx}
\usepackage{color}
\usepackage{tikz}
\usepackage{psfrag}
\usepackage{soul}

\setcitestyle{square,numbers}

\begin{document}

\newcommand{\IM}{{\rm Im}\,}
\newcommand{\card}{\#}
\newcommand{\la}[1]{\label{#1}}
\newcommand{\eq}[1]{(\ref{#1}),} 
\newcommand{\figref}[1]{Fig. \ref{#1}}
\newcommand{\abs}[1]{\left|#1\right|}
\newcommand{\comD}[1]{{\color{red}#1\color{black}}}
\newcommand{\p}{{\partial}}
\newcommand{\Tr}{{\text{Tr}}}
\newcommand{\tr}{{\text{tr}}}
\newcommand{\sym}{${\cal N}=4$ SYM becomes a non-unitary and non-supersymmetric CFT. }
\newcommand{\como}[1]{{\color[rgb]{0.0,0.1,0.9} {\bf \"O:} #1} }
\newcommand{\modd}[1]{\textcolor{red}{#1}}

\makeatletter
\newcommand{\subalign}[1]{%
  \vcenter{%
    \Let@ \restore@math@cr \default@tag
    \baselineskip\fontdimen10 \scriptfont\tw@
    \advance\baselineskip\fontdimen12 \scriptfont\tw@
    \lineskip\thr@@\fontdimen8 \scriptfont\thr@@
    \lineskiplimit\lineskip
    \ialign{\hfil$\m@th\scriptstyle##$&$\m@th\scriptstyle{}##$\crcr
      #1\crcr
    }%
  }
}
\makeatother

\newcommand{\mzvv}[2]{
  \zeta_{
    \subalign{
      &#1,\\
      &#2
    }
}
}

\newcommand{\mzvvv}[3]{
  \zeta_{
    \subalign{
      &#1,\\
      &#2,\\
      &#3
    }
}
  }

\makeatletter
     \@ifundefined{usebibtex}{\newcommand{\ifbibtexelse}[2]{#2}} {\newcommand{\ifbibtexelse}[2]{#1}}
\makeatother

\preprint{LPTENS--18/01, ZMP--HH--18/04}


\usetikzlibrary{decorations.pathmorphing}
\usetikzlibrary{decorations.markings}
\usetikzlibrary{intersections}
\usetikzlibrary{calc}

\tikzset{
photon/.style={decorate, decoration={snake}},
particle/.style={postaction={decorate},
    decoration={markings,mark=at position .5 with {\arrow{>}}}},
antiparticle/.style={postaction={decorate},
    decoration={markings,mark=at position .5 with {\arrow{<}}}},
gluon/.style={decorate, decoration={coil,amplitude=2pt, segment length=4pt},color=purple},
wilson/.style={color=blue, thick},
scalarZ/.style={postaction={decorate},decoration={markings, mark=at position .5 with{\arrow[scale=1]{stealth}}}},
scalarX/.style={postaction={decorate}, dashed, dash pattern = on 4pt off 2pt, dash phase = 2pt, decoration={markings, mark=at position .53 with{\arrow[scale=1]{stealth}}}},
scalarZw/.style={postaction={decorate},decoration={markings, mark=at position .75 with{\arrow[scale=1]{stealth}}}},
scalarXw/.style={postaction={decorate}, dashed, dash pattern = on 4pt off 2pt, dash phase = 2pt, decoration={markings, mark=at position .60 with{\arrow[scale=1]{stealth}}}}
}

 \newcommand{\doublewheelsmall}{
   \begin{minipage}[c]{1cm}
     
     \begin{center}
       \begin{tikzpicture}[scale=0.3]
         \foreach \m in {1,2} {
           \draw (0.75*\m,0) arc[radius = 0.75*\m,start angle = 0, end angle = 300] ;
           \draw[black, densely dotted] (300:0.78*\m) arc[radius = 0.75*\m,start angle = -60, end angle = 0];
         }
         \foreach \t in {1,2,...,5} {
           \draw (0,0) -- (60*\t:1.5);
         }
         \draw[black,densely dotted] (0,0) -- (0:1.5);
       \end{tikzpicture}
     \end{center}
   \end{minipage}
 }


\newcommand{\footnoteab}[2]{\ifbibtexelse{%
\footnotetext{#1}%
\footnotetext{#2}%
\cite{Note1,Note2}%
}{%
\newcommand{\textfootnotea}{#1}%
\newcommand{\textfootnoteab}{#2}%
\cite{thefootnotea,thefootnoteab}}}

\def\e{\epsilon}
     \def\bT{{\bf T}}
    \def\bQ{{\bf Q}}
    \def\wT{{\mathbb{T}}}
    \def\wQ{{\mathbb{Q}}}
    \def\ttQ{{\bar Q}}
    \def\tQ{{\tilde \bP}}
        \def\bP{{\bf P}}
        \def\dq{{\dot q}}
    \def\CF{{\cal F}}
    \def\cC{\CF}
    
     \def\l{\lambda}
\def\hbZ{{\widehat{ Z}}}
\def\bZ{{\resizebox{0.28cm}{0.33cm}{$\hspace{0.03cm}\check {\hspace{-0.03cm}\resizebox{0.14cm}{0.18cm}{$Z$}}$}}}

\title{ Bi-scalar integrable CFT at any dimension }

\author{Vladimir Kazakov$^{a,b}$ and\, Enrico Olivucci$^{c}$}

\affiliation{%
\\
\(^{a}\) Laboratoire de Physique Th\'{e}orique de l'\'{E}cole Normale Sup\'{e}rieure, 24 rue Lhomond,
F-75231 Paris Cedex 05, France  
 \\
\(^{b}\) PSL University, CNRS, Sorbonne Universit\'{e}s, UPMC Universit\'{e} Paris 6  \\
\(^{c}\) 
II. Institut f\"ur Theoretische Physik, Universit\"at Hamburg, Luruper Chaussee 149, 22761
Hamburg, Germany
\\
}

\begin{abstract}
We propose a \(D\)-dimensional generalization of \(4D\) bi-scalar conformal quantum field theory recently introduced by G\"urdogan and one of the authors as a {particular strong-twist limit of \(\gamma\)-deformed} \({\cal\ N}=4\) SYM theory. {Similarly to the \(4D\) case, the planar correlators of this \(D\)-dimensional theory are conformal and dominated by ``fishnet" Feynman graphs. The dynamics of these graphs is described by the integrable conformal \(SO(1,D+1)\) spin chain.} In 2\(D\) it is the analogue of L.~Lipatov's \(SL(2,C)\) spin chain for the Regge limit of \(QCD\), but with the spins \(s=1/4\) instead of \(s=0\). Generalizing recent  \(4D\) results of Grabner, Gromov, Korchemsky and one of the authors  to any \(D\)  we compute exactly, at any coupling, a four point correlation function, dominated by the simplest fishnet graphs of cylindric topology, and extract from it exact dimensions {of operators with chiral charge 2 and any spin, together with some of their Operator Product Expansion structure constants.}
\begin{center}
\textit{the paper is dedicated to the memory of L.N.~Lipatov}  
\end{center}
\end{abstract}  

 \maketitle

\section{Introduction}

Conformal field theories (CFT) are ubiquitous in two dimensions \cite{DiFrancesco1997}, and    quite a few supersymmetric  CFTs in \(D=3,4,6\) dimensions are known.
But well defined non-supersymmetric CFTs in \(D>2\), such as 3D Ising or Potts models,   or Banks-Zaks model \cite{Banks1982b}, are rare species, {in spite of their rich potential applications ranging from the theory of phase transitions to fundamental interactions}.  The CFTs at \(D>2\) which in addition are integrable, such as 4D \({\cal\ N}=4\) SYM and Aharony-Bergman-Jafferis-Maldacena (ABJM) theories in 't~Hooft limit, are true exceptions \cite{Beisert:2010jr}~\footnote{The integrability of such theories emerges due to their duality to string sigma  models on specific cosets, having an infinite  number of quantum conservation lows for their world-sheet dynamics, or due to the analogy between the planar Feynman diagram technique and integrable \((1+1)D\) quantum spin chains~\cite{Beisert:2010jr}.}. That's why a new family of {planar} integrable CFTs obtained in  \cite{Gurdogan:2015csr} as a special double scaling limit of \(\gamma\)-deformed \({\cal\ N}=4\) SYM seems to be an important and instructive example. This theory can be studied via quantum spectral curve (QSC) formalism \cite{Gromov2014a,Gromov:2009tv,Kazakov:2015efa} or using the integrability of its dominant   Feynman graphs via the conformal, \(SU(2,2)\) noncompact spin chain. A nice particular  case of this family is the \(4D\) bi-scalar theory, { whose planar limit is} dominated by "fishnet" type Feynman graphs \cite{Gurdogan:2015csr,Caetano:2016ydc}.

We propose here the following \(D\)-dimensional generalization of the \(4D\) bi-scalar theory  introduced in \cite{Gurdogan:2015csr}
\begin{align}
    \label{bi-scalarL-alpha}
    {\cal L}_{\phi}&=  N_c\,\tr
    [\phi_1^\dagger \,\, (-\p_\mu \p^\mu)^{\omega}\,\phi_1 + \phi_2^\dagger \,\, (-\p_\mu \p^\mu)^{\frac{D}{2}-\omega}\,\phi_2 \notag
\\
&+(4 \pi)^\frac{D}{2} \xi^2\phi_1^\dagger \phi_2^\dagger \phi_1\phi_2].
  \end{align}
  {where both scalar fields transform under the adjoint representation of \(SU(N_c)\); \(\xi^2\) is the coupling constant and \(\omega\,\in\,\left(0,\frac{D}{2}\right)\) is a deformation parameter.}
{The }non-local (for general \(D,\,\omega\)) operators in kinetic terms should be understood as an integral kernel
\begin{align}
    \label{propagator}
    (\p_\mu \p^\mu)^{\beta}f(x)\equiv \frac{(-4)^\beta \,\Gamma(\frac{D}{2}+\beta)}{\pi^\frac{D}{2} \Gamma(-\beta)}\int\frac{d^D y\,f(y)}{|x-y|^{{D}+2\beta}}.
  \end{align}The propagator of scalar fields is its functional inverse:
\begin{align}
    \label{bi-scalarL}
     (-\p_\mu \p^\mu)^{\beta} D(x)&= \,\delta^{(D)}(x),
     \\ \notag D(x-y)&= \frac{\Gamma(\frac{D}{2}-\beta)}{4^\beta \,\pi^\frac{D}{2} \Gamma(\beta) \,\,|x-y|^{D-2\beta}}.
  \end{align}
The typical structure  in the bulk of sufficiently big planar Feynman graphs in this  theory is that of the regular square lattice (``fishnet" graphs, proposed in ~\cite{Zamolodchikov:1980mb} as an integrable lattice spin model), by the same reasons as in \(4D\) case  \cite{Gurdogan:2015csr,Caetano:2016ydc}, namely, due to the presence of the single chiral interaction vertex in the Lagrangian, and the absence of its hermitian  conjugate.
 For example, the graphs renormalizing local ``vacuum" operator \(\tr(\phi_j)^L\) are those of the ``wheel" type and they can be studied via the integrable conformal {\(SO(2,D)\)} spin chain~\footnote{in the rest of the paper we will use its Euclidean version \(SO(1,D+1)\) instead of the Minkowskian \(SO(2,D)\).}, as was suggested for \(4D\) case in \cite{Gromov:2017cja}. {The dimensions of operators of the type \(\tr[\phi_1^3(\phi_2^\dagger\phi_2)^k]\) have been also studied in \(4D\) \cite{Gromov:2017cja} by QSC methods. It is not clear whether this method can be generalized to our \(D\) dimensional model. But the spin chain methods certainly can.} 

In general, the propagators of the fishnet graphs of the model \eqref{bi-scalarL-alpha} are   different in two different directions: \(|x-y|^{-D+2\omega}\) for \(\phi_1\) fields and \(|x-y|^{-2\omega}\) for \(\phi_2\) fields. Let us concentrate here on the ``isotropic" case  \(\omega=D/4\). In order to maintain  the renormalizability we should add to \eqref{bi-scalarL-alpha} the following double-trace counterterms \cite{Fokken:2013aea,Fokken:2014soa}
\begin{align}
\label{double-tr-L}
{}& {\cal L}_{\rm dt}/(4\pi)^\frac{D}{2}=\alpha_1^2 \, \sum_{i=1}^2\tr(\phi_i\phi_i)\,\tr(\phi_i^{\dagger}\phi_i^{\dagger})     - 
 \notag \\ 
{}& - \alpha_2^2\,\tr(\phi_1\phi_2)\tr(\phi_2^{\dagger }\phi_1^{\dagger })
-\alpha_2^2\tr(\phi_1\phi_2^{\dagger })\tr(\phi_2\phi_1^{\dagger })\, ,
\end{align}
%
%
  Notice that the first term disappears in the ``non-isotropic" case \(\omega\ne \frac{D}{4}  \) since the couplings of two terms in the first line of \eqref{double-tr-L} would become dimensionful.

\begin{figure}
  \includegraphics[scale=0.55]{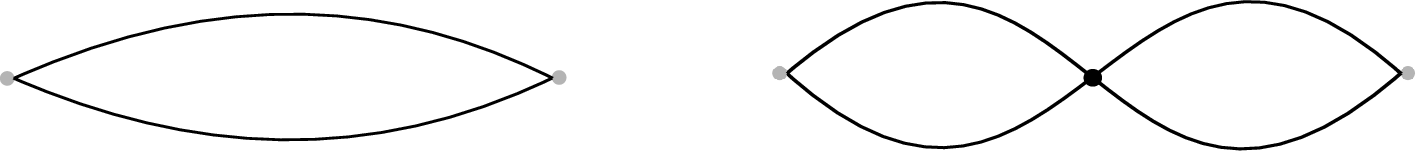}\vspace*{20pt}
  \includegraphics[scale=0.55]{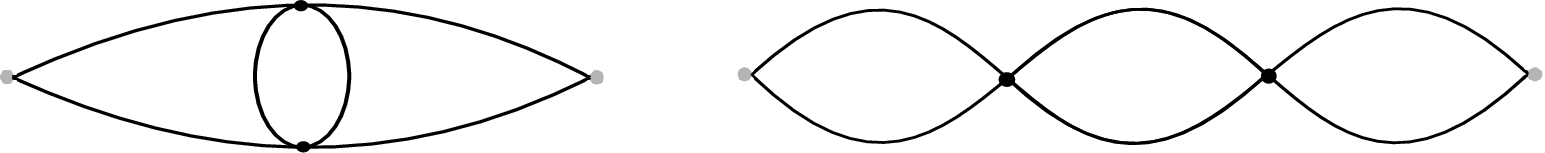}
  \caption{Loop expansion of \(\langle\tr(\phi_1^2)(x)\tr(\phi_1^2)^\dagger(0) \rangle\) {planar} graphs up to 2-loops.}
\label{2points}
\end{figure}

 As it was suggested  in   \cite{Sieg:2016vap} and explicitly shown in \cite{Grabner:2017pgm} for the 4D case, the ``isotropic" bi-scalar theory with Lagrangian \({\cal L}_{\phi}+{\cal L}_{\rm dt}$   has  two fixed points.
We generalize here this result to any dimension, up to two loops, computing the corresponding Feynman graphs (Fig.\ref{2points}) contributing to  the \(\beta_{\alpha_1}\)-function. Its two zeroes are \begin{equation}
 \alpha_{1}^2 (\xi)= \mp \frac{i\, \xi ^2}{2}  - J^{(D)} {\xi^4} + O(\xi^6) \label{critic}
\end{equation}   
where the real coefficient \(J^{(D)}\) depends on the \(\epsilon^{-1}\) coefficient of the down-left graph of Fig.\ref{2points} in dimensional regularization. For example: \({J}^{(4)}=1/2\),\(\;{J}^{(2)}=2 \ln 2\), \( J^{(1)}=\frac{\pi+4 \ln 2}{2\sqrt{\pi}}\) \footnote{This term can be computed at any D by means of Integration by Parts and Mellin-Barnes transformation, according to \cite{Grozin2012}}. 
  At this critical coupling \(\alpha_1(\xi)\) the bi-scalar theory becomes a genuine non-unitary CFT at any coupling \(\xi\).  
 The operators $\tr(\phi_1\phi_2)$,
 and $\tr(\phi_1\phi_2^{\dagger})$  are protected in the planar limit {as in} ~\cite{Grabner:2017pgm}. 
 
 In this paper, generalizing the \(4D\) results of ~\cite{Grabner:2017pgm} to any \(D\), we will compute exactly a particular  four-point function  and read off from it the exact scaling dimensions and certain OPE structure constants of   operators of the type  \(\tr(\phi_1\partial_+^S\phi_1(\phi_2^\dagger\phi_2)^k)+\text{permutations}.\,\, \) Their dimensions will be  given by
a remarkably simple exact relation
 \begin{align} \label{ex-eq}
h_{\Delta,S}\equiv \frac{   \Gamma \left(\frac{3D}{4}-\frac{\Delta- S}{2}\right)  }{\Gamma\left(\frac{D}{4}-\frac{\Delta- S}{2}\right)  }\frac{   \Gamma \left(\frac{D}{4}+\frac{\Delta+ S}{2}\right)  }{\Gamma\left(-\frac{D}{4}+\frac{\Delta+ S}{2}\right)  }=  \xi^4\,,
\end{align} which reduces of course at  \(4D\) to the result of ~\cite{Grabner:2017pgm}.
For even \(D\) it gives \(D\) different solutions  $\Delta(\xi)=\Delta_0+\gamma(\xi)$. At odd (or non-integer) \(D\) there are infinitely many, in general complex, solutions. At weak coupling the two complex conjugate solutions at $S=0$ { \footnote{Here and in the following we adopt the notation \(\psi^{(k-1)}(x)\,=\,\frac{d^k}{d x^k} \, \log \Gamma(x)\) for polygamma functions.}} \begin{align*}\gamma=\pm i \frac{2\xi ^2}{\Gamma \left(\frac{D}{2}\right)}\pm\frac{i}{6} \frac{\xi^6}{\Gamma\left(\frac{D}{2}\right)^3}\left(\pi ^2-6\, \psi
   ^{(1)}\left(\frac{D}{2}\right)\right) +O(\xi^{10})\end{align*} describe anomalous dimensions of the operator $\tr(\phi_1\phi_1)$ at the two
fixed points. In a similar way, for any $S \in 2\mathbb{Z}$ the real weak coupling solution\begin{align*}&\gamma=-2 \frac{ \xi ^4 \Gamma (S)}{\Gamma
   \left(\frac{D}{2}\right) \Gamma \left(\frac{D}{2}+S\right)}+ \frac{2\, \xi ^8 \Gamma (S)^2}{\Gamma \left(\frac{D}{2}\right)^2 \Gamma \left(\frac{D}{2}+S\right)^2} \times \\& \times\left(\psi ^{(0)}\left(\frac{D}{2}\right)-\psi ^{(0)}\left(\frac{D}{2}+S\right)+\psi ^{(0)}(S)+{\gamma_E}
   \right) +O(\xi^{10}).
\end{align*}
describes the operators of the type \(\tr(\phi_1\partial_+^S\phi_1)\), {where \(\partial_+^S = (\hat n\cdot \partial)^S\) with \(\hat n\) being an auxiliary light-like vector.}\\
For \(D=2m,\,\,m\in \mathbb{N}\) the L.H.S. of \eqref{ex-eq} factorizes into a polynomial of  degree \(2m\) and \(2m\) roots of eq.\eqref{ex-eq} describe the scaling dimension of the exchanged operators in the OPE channel \(x_3\rightarrow x_4\) of \eqref{G4} together with their shadows \(\tilde \Delta = D-\Delta \).
At \(\xi=0\) we get for the bare dimensions of physical operators (i.e., excluding ``shadow" operators) \begin{align*}
\Delta_0-S = \{m,m+2,\cdots, 3m-2\},
\end{align*}
At \(D=2\) there is a single solution with the dimension \(\Delta= 1+\sqrt{S^2-4\xi^4}\) of  the local twist-2 operators of the type \(\tr(\phi_1 \partial_+^S\phi_1)\), while at \(4D\) the additional $\Delta_0-S = 4$ describes twist-4 operators~\cite{Grabner:2017pgm}.\\As an example, at \(D=6\) and \(S=0\) the possible non-shadow solutions for \eqref{ex-eq} are   \(\Delta_0=3,5,7\). They  can be realized as \(\tr(\phi_1^2)\) for \(\Delta_0=3\),   linear combinations of \( \tr(\phi_1\Delta \phi_1)\), \(\tr(\p_\mu \phi_1 \p^\mu \phi_1)\) for \(\Delta_0=5\) and of \(\tr(\phi_1\Delta ^2\phi_1)\), \(\tr(\Delta\phi_1\Delta \phi_1)\), \(\tr(\p_\mu\phi_1 \p^\mu\Delta ^2\phi_1)\), \(\tr(\p_\mu \p_\nu \phi_1\p^\mu \p^\nu \phi_1)\) for \(\Delta_0=7\).  

 Diagonalizing the mixing matrix of these operators at \(\xi\ne0\) we would obtain operators with non-trivial, \(\xi\)-dependent anomalous dimensions, as well as the so called log-multiplets~\cite{Caetano:TBP,Gromov:2017cja},     omnipresent in this non-unitary theory \cite{Gurarie:1993xq,Hogervorst:2016itc}, containing the operators with zero anomalous dimension.    Eq.\eqref{ex-eq} predicts that all the exchange operators  from this set acquire non-trivial anomalous dimensions,   whereas the operators belonging to log-multiplets never appear among them. This appears to be true at any even dimension \(D\). \\  

As a general rule, according to the eq.\eqref{ex-eq}  the operators of the type \(\lbrace\tr(\phi_1\partial_+^S \phi_1 (\phi_2\phi_2^\dagger)^k)+\text{permutations}\rbrace\) appear in the multiplets only at \(D/4\,\in\,\mathbb{N}\), \(k=1\). We will  find below from the exact 4-point function the conformal structure constants of these operators with two scalar fields.
 
\section{Integrability of \(D\)-dimensional bi-scalar CFT}

As it was noticed in  \cite{Gurdogan:2015csr}  and further  developed in \cite{Caetano:2016ydc,Gromov:2017cja,Grabner:2017pgm}, the \(4D\) case of the theory
\eqref{bi-scalarL-alpha}, with \(\omega=1\), is integrable in the planar limit. On the one hand, this integrability is the direct consequence of integrability of \(\gamma\)-twisted  planar  \({\cal\ N}=4\) SYM theory, from which it was obtained in the double scaling  limit combining strong imaginary twist  and weak coupling. On the other hand, this integrability was explicitly related in   \cite{Gurdogan:2015csr,Gromov:2017cja}  to the fact that the bi-scalar theory was dominated by the integrable ``fishnet" Feynman graphs~\cite{Zamolodchikov:1980mb},\footnote{this terminology for the graphs of regular square lattice shape was introduced by  B.Sakita and M.Virasoro in 1970}. 

\begin{figure}
\includegraphics[scale=0.29]{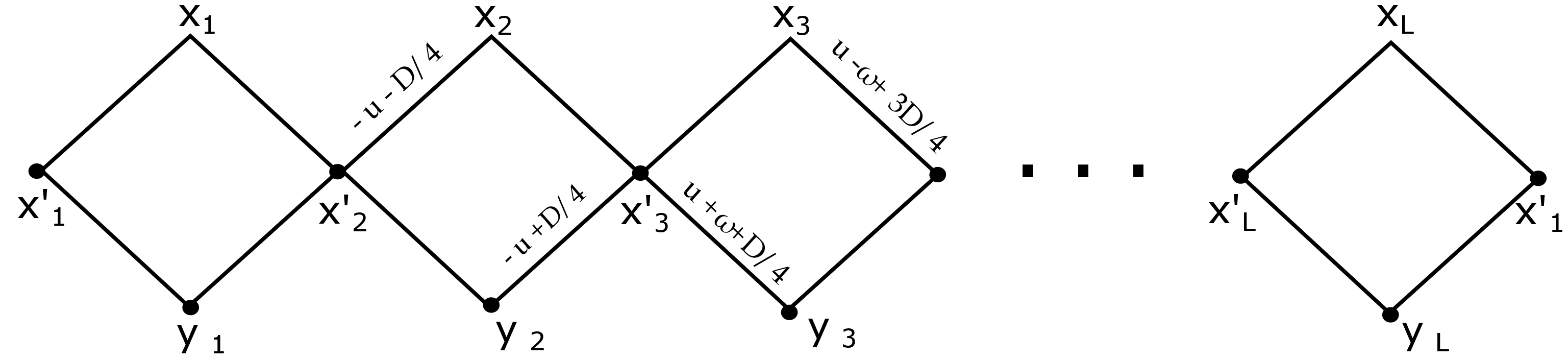}
\caption{Graphical representation of the transfer matrix as a convolution of R-kernels according to formulas \eqref{transfer_matrix}and \eqref{Rmat}. Black dots are integration points and the weights of propagators are written in the second and third R-kernel.}
\label{t_matrix}
\end{figure}

\begin{figure}
\includegraphics[scale=0.26]{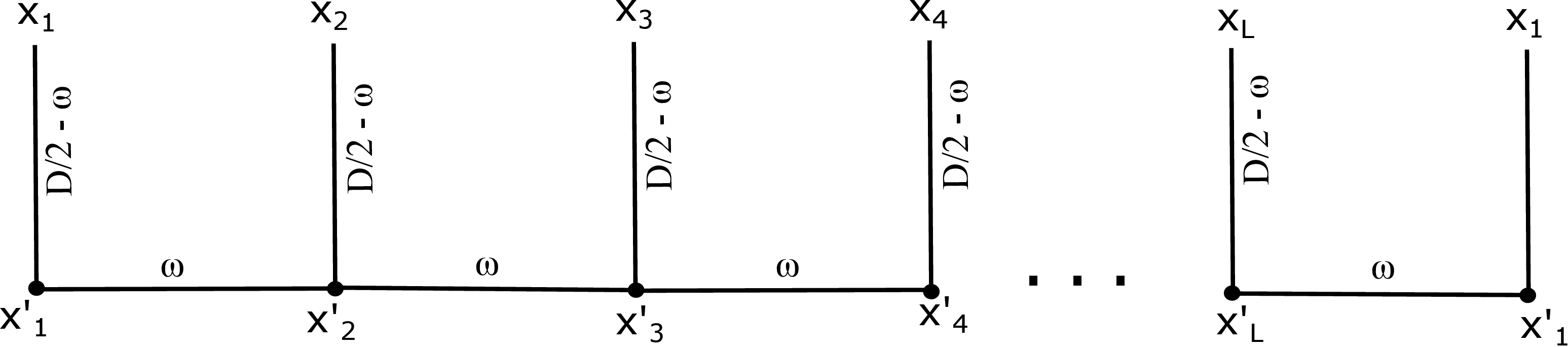}
\caption{Graphical representation of the kernel of the graph-building operator for generic \(D\) and \(\omega\). It is otained by setting \(u=-\frac{D}{4}\) in the transfer matrix \eqref{transfer_matrix} presented on Fig.~\ref{t_matrix}, so that \(x_{jj'+1~}\)--type type propagators disappear while \(x_{j'+1}-y_{j}\)~-type propagators are replaced by \(\delta^{(D)}(x_{j'+1}-y_{j})\) factors. After that, integration over the points \(y_{j}\)  is equivalent to setting \(y_{j}=x_{j'+1}\).}
\label{comb}
\end{figure}

Apart from \(4D\) case, at arbitrary \(D\)  our bi-scalar model \eqref{bi-scalarL-alpha} does not have any integrable SYM origin. But the arguments of equivalence to the integrable conformal \(SO(1,D+1)\) spin chain do work.   Namely,   let us introduce the \(D\)-dimensional analogue of the \(4D\)  "graph-building" operator    \cite{Gurdogan:2015csr} {at general \(\omega\)-deformation} \begin{align} \label{graph-building} & \mathcal H \, _L\Phi(x_1,\dots ,x_L) =\notag\\
&\frac{1}{\pi^\frac{D L}{2}}\int {d^D x_{1'} \dots d^D x_{L'}\,\Phi(x_{1'},\dots,x_{L'})\over |x_{11'}|^{D-2\omega}\dots |x_{_{LL'}}|^{D-2\omega}  \times|x_{1'2'}|^{2\omega}\dots |x_{L'1'}|^{2\omega}}
\end{align}  schematically presented on Fig.\ref{comb}. It is easy to see that a power of this operator \(\mathcal{H}_L^M\) generates a fishnet Feynman graph with topology of a cylinder of length \(M\) with the circumference \(L\). Now, in analogy with the \(4D\) observation of \cite{Gromov:2017cja}, we notice that this operator can be related to the transfer-matrix of integrable \(SO(1,D+1)\) conformal Heisenberg spin chain~\cite{Chicherin2013a} presented on   Fig.\ref{t_matrix}: \begin{align}\label{transfer_matrix}
\mathbb{T}(u)=\Tr_0\left(R_{01}(u)\, R_{02}(u)\dots R_{0L}(u)\right)
\end{align} 
 {where \(u\) is the spectral parameter  and the \(R\)-matrix acts as an integral operator}
\begin{align} & \mathcal [R \, _{12}\Phi](x_1,x_2)(u) = c(u,D,\omega)\,\times \notag\\
\times&\int {d^D x _{1'}d^D x_{2'}\,\Phi(x_{1'},x_{2'})\over (x_{12}^2)^{-u-\frac{D}{4}} (x_{21'}^2)^{\frac{D}{4}+u + \omega}(x_{12'}^2)^{\frac{3D}{4}+u - \omega}(x_{1'2'}^2)^{-u+\frac{D}{4}} }, \label{Rmat} \end{align}  \text{with the normalization constant }
\begin{align*}c(u,D,\omega)=\frac{4^{2u}}{\pi^D}\frac{\Gamma\left(u+\frac{D}{4}+\omega\right)\Gamma\left(u+\frac{3D}{4}-\omega\right)}{\Gamma\left(-u-\frac{D}{4}+\omega\right)\Gamma\left(-u+\frac{D}{4}-\omega\right)}.
\end{align*}
Indeed, in analogy with \(4D\) case~\cite{Gromov:2017cja},  at a particular value of spectral parameter this transfer matrix becomes the graph-building operator \eqref{graph-building} at any \(D\)
\begin{align}\mathcal{H}_L=\pi^{-\frac{D L}{2}}\left[{(4\pi^2)^\frac{D}{2}}\Gamma\left(\frac{D}{2}\right)\right]^L\,\lim_{\epsilon \rightarrow 0}\,\epsilon^{L}\,\mathbb{T}\left(-\frac{D}{4}+\epsilon\right).
\end{align}     presented on Fig.~\ref{comb}.
{Thus this operator is one of the conserved charges of the equivalent spin chain:  \(\mathbb{[T}(u),\mathbb{T}(u')]=\mathbb{[T}(u),\mathcal{H}_L]=0\).}\\ 
\section{Exact 4-points correlation function}

In analogy with \(4D\) results of ~ \cite{Grabner:2017pgm}, employing the  \(D\)-dimensional   conformal symmetry of the theory \eqref{bi-scalarL-alpha},\eqref{double-tr-L} we will compute exactly the four-point correlation function 
\begin{align}\label{G4}
G= \langle O(x_1,x_2) \bar O (x_3,x_4) \rangle = { \mathcal G(u,v) \over (2\pi)^D (x_{12}^2 x_{34}^2)^\frac{D}{4}} \,,
\end{align}
where the notation is introduced for the operators \(O(x,y)= \tr[\phi_1 (x) \phi_1(y)]\) and \(\bar O(x,y)=\tr[\phi_1^\dagger (x)\phi_1^\dagger(y)]\).\\
Here \(\mathcal G(u,v)\) is a finite function of cross-ratios \(u=x_{12}^2 x_{34}^2/(x_{13}^2 x_{24}^2)\)
and \(v=x_{14}^2x_{23}^2/(x_{13}^2 x_{24}^2)\), invariant under the exchange of points \(x_1\leftrightarrow x_2\) and
\(x_3\leftrightarrow x_4\). The OPE expansion leads to the  formula 
\begin{align}\label{cpwe}
\mathcal G(u,v) =\sum_{\Delta} \sum_{ S/2\in \mathbb Z_+}C^2_{\Delta, S}\, u^{(\Delta-S)/2} g_{\Delta,S} (u,v),
\end{align}
where the sums run over operators with scaling dimensions \(\Delta\) and even Lorentz spin $S\). Here
 \(C_{\Delta, S}\) is the corresponding OPE coefficient (structure constant) and \(g_{\Delta,S} (u,v)\) is the known  \(D\) dimensional conformal block (see (2.9) and sections 4,5 in \cite{Dolan2011}).
If we  compute 
\eq{G4} we will  identify the conformal data for the operators emerging in the OPE of \(O(x_1,x_2) \).  

\begin{figure}
 \includegraphics[scale=0.6]{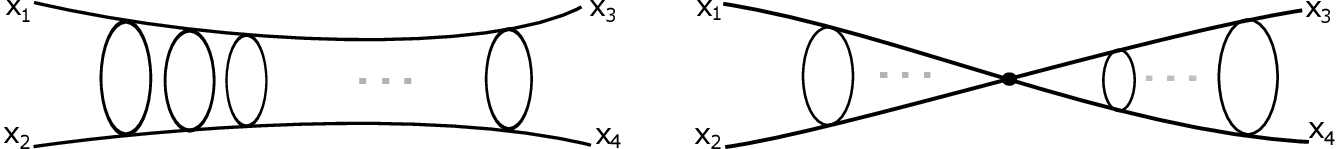}\vspace*{20pt}
  \includegraphics[scale=0.6]{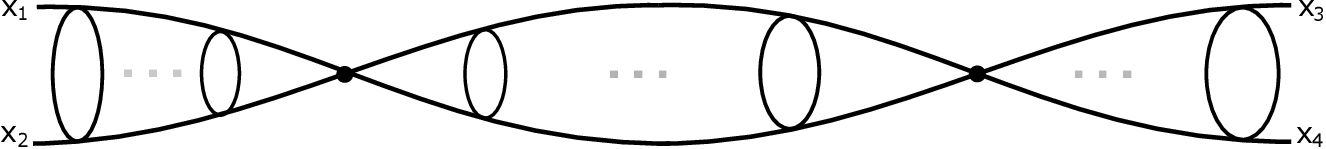}
  \caption{General fishnet graphs up to \(\alpha^2_1\) order in the expansion of four point function \eqref{G4-int}.}
\label{fig}
\end{figure}

In the planar limit \(G\) is given by the  set of fishnet Feynman diagrams presented in Fig.~\ref{fig}. 
Summing up the corresponding perturbation series we encounter a geometric progression involving the combination of operators \(\alpha^2 \mathcal V+ \xi^4 \mathcal H_2   \), where  $\alpha^2=\alpha_\pm^2$ is the double-trace coupling at the fixed point,   $\mathcal V$ is the operator inserting the double-trace vertex\begin{align*}
& \mathcal V ~\Phi(x_1,x_2) = \frac{2}{\pi^{\frac{D}{2}}}\int {d^D x_{1'} d^D x_{2'}\,\delta^{(D)}(x_{1'2'})\,\Phi(x_{1'},x_{2'})\over |x_{11'}|^{D/2}|x_{22'}|^{D/2}},
\end{align*}which is the  D dimensional version of (11) in~\cite{Grabner:2017pgm}, and the operator
\(\mathcal H_2\) defined by \eqref{graph-building} adds a scalar loop inside  the
diagram. Hence we obtain the following  representation 
\begin{align}\label{G4-int} 
G=\frac{1}{(2\pi)^D}\int {d^4 x_{3'} d^4 x_{4'}\over \left(x_{33'}^2 x_{44'}^2\right)^\frac{D}{4}}
\langle x_{1},x_{2}| {1\over 1-\alpha^2 \mathcal V- \xi^4 \mathcal H_2} |x_3',x_4'\rangle+\notag\\+(x_1\leftrightarrow x_2).
\end{align}
where \(x_{ij}\equiv x_i-x_j\).~\footnote{The operators $\mathcal V$ and $\mathcal H$ are not well-defined separately, e.g.
for an arbitrary \(\Phi(x_i)\) the expressions for \(\alpha^4 \mathcal V^2 \Phi(x_i)\) and \(\xi^4 \mathcal H  \Phi(x_i)\) are given by  
divergent integrals. However, at the fixed point, their sum is finite by virtue of conformal symmetry.}

Remarkably,  the operators $ \mathcal V$ and $ \mathcal H_2$  commute  with the generators of the conformal group, as in the particular \(4D\) case~\cite{Grabner:2017pgm}.  This 
fixes the form of their eigenstates 
\begin{align}\label{Phi-def}
\Phi_{\Delta,S,n}(x_{10},x_{20}) =  {1\over \left(x_{12}^2\right)^\frac{D}{4}} \left(x_{12}^2\over x_{10}^2 x_{20}^2\right)^{(\Delta-S)/2}\!\!
{\left(\partial_{0} \ln {x_{20}^2\over x_{10}^2}\right)^S},
\end{align}
where {$\Delta=\frac{D}{2}+2i\nu$ and $\partial_0 \equiv (\hat n \cdot \partial_{x_0})$}. The state \(\Phi_{\Delta,S,n}\) belongs to the principal series of the conformal group and can be  represented in the form of a conformal three-point correlation function
\begin{align*}
C_{\Delta,S} \,\Phi_{\Delta,S,n}(x_{10},x_{20}) = \langle\,\text{tr}[\phi_1(x_1) \phi_1(x_2)] O_{\Delta,S,n}(x_0)\,\rangle\,,
\end{align*}
where the operator $O_{\Delta,S,n}(x_0)$ carries the scaling dimension $\Delta$ and Lorentz spin $S$, and \(C_{\Delta,S}\) is the 3-points structure constant.
The states \eq{Phi-def} satisfy the orthogonality condition~\cite{Tod:1977harm, Fradkin1978}
 \begin{align}\label{states}\notag
& \int   \frac{d^Dx_{1}d^Dx_{2}}{(x_{12}^2)^\frac{D}{2}}\,\overline{ \Phi_{\Delta',S',n'}}(x_{10'},x_{20'}) \,\Phi_{\Delta,S,n}(x_{10},x_{20})     
\\\notag
&= 
 c_1(\nu,S) \delta(\nu-\nu')\,\delta_{S,S'}\delta^{(D)}(x_{00'})(nn')^S
 \\
 &+ 
 c_2(\nu,S) {\delta(\nu+\nu')\delta_{S,S'}} Y^S(x_{00'}) /(x_{00'}^2)^{{\frac{D}{2}-S-2i\nu}},
\end{align} 
where $\Delta'=\frac{D}{2}+2i\nu'$, $Y(x_{00'})\!=\!(n \p_{x_0} )(n'\p_{x_{0'}})\ln x_{00'}^2 $, and
\begin{align}
c_1(\nu,S)=& \frac{ 2^{{S}+1}\, S!\,\left|\Gamma(2 i \nu)\right|^2  \left( 4 \nu ^2
   +(\frac{D}{2}+S-1)^2\right)^{-1}}{\pi ^{-(3D /2  + 1)} \left|\Gamma\left(\frac{D}{2}-1+2i\nu\right) \right|^2 \Gamma(\frac{D}{2}+S)},
 \\ \notag
 c_2(\nu,S)=&\frac{2 (-1)^S\, \Gamma^2\! \left(\frac{D}{4}+\frac{S}{2}-i \nu\right)  }{ \pi ^{-(D+1)}\, \Gamma^2 \!\left(\frac{D}{4}+\frac{S}{2}+i \nu\right) }\frac{\Gamma(2i\nu)}{\Gamma(\frac{D}{2}+2 i \nu -1)}\\ \notag
 &\times \frac{\Gamma (\frac{D}{2}+S+2 i \nu-1)\,S! }{ \Gamma (\frac{D}{2}+S-2 i \nu ) \Gamma(\frac{D}{2}+S)}
\end{align}
 Calculating the corresponding eigenvalues of the operators $ \mathcal V$ and $ \mathcal H$ we find
\begin{align}\notag \label{eigen}
{}& \mathcal V\, \Phi_{\Delta,S,n}(x_1,x_2)  = \delta(\nu) \delta_{S,0}  \Phi_{\Delta,S,n}(x_1,x_2),
\\
{}& \mathcal H \,\Phi_{\Delta,S,n}(x_1,x_2)  =   h^{-1}_{\Delta,S} \Phi_{\Delta,S,n}(x_1,x_2),
\end{align}
where the function $h(\Delta,S)$ is given by \eqref{ex-eq}.
Applying \eqref{states}--\eq{eigen} we can expand the correlation function \eqref{G4-int} over the basis of states \eqref{Phi-def}.
This yields the expansion of $G$ over conformal partial waves defined by the operators $O_{\Delta,S}(x_0)$ in the OPE channel
\(O(x_1,x_2)\)
\begin{align}\label{G-cont}
\mathcal G(u,v)=\sum_{S/2\in \mathbb Z_+} \int_{-\infty}^\infty d\nu\mu_{\Delta,S}\,
{u^{(\Delta-S)/2} g_{\Delta,S} (u,v)
\over h_{\Delta,S}- \xi^4},
\end{align}
where $\Delta=\frac{D}{2}+2i\nu$, and $\mu_{\Delta,S}=2 \pi^D/c_2(\nu,S)$ is related to the norm of the state \eqref{states}. The fact that the dependence on $\alpha^2$ disappears from \eq{G-cont} can be understood as follows. Viewed as a function of $S$, $\xi^4/h_{\Delta,S}$ develops  poles at $\nu = \pm i S$ which pinch the integration contour in \eq{G-cont} for $S\to 0$.  The contribution of the operator 
$\mathcal V$ is needed to make a perturbative expansion of \eqref{G-cont} well-defined. 
For finite $\xi^4$, these poles provide a vanishing contribution to \eq{G-cont} but generate a branch-cut $\sqrt{-\xi^4}$ singularity of $\mathcal G(u,v)$, as in \(4D\) case~\cite{Grabner:2017pgm}.
 
At small $u$, we close the integration
contour in \eqref{G-cont} to the lower half-plane and pick up residues at the poles located at solutions of   
\eqref{ex-eq}
and satisfying the unitarity bound ${\rm Re}\,\Delta>S$. The resulting expression for $\mathcal G(u,v)$ takes the expected form
\eq{cpwe} with the OPE coefficients given by 
\begin{align}\notag
C^2_{\Delta, S} {}& = \frac{\, \Gamma(\frac{D}{2}+S)}{S!\,}\frac{\Gamma(\Delta -1)}{\Gamma(\Delta-\frac{D}{2})}\text{Res}\left(\frac{d\Delta}{h_{\Delta,S}-\xi^4}\right) 
   \\ \label{ex2}
{}& \times   \frac{\Gamma (S-\Delta +D) \,\Gamma ^2\left(\frac{1
   }{2}(S+\Delta)\right)}{ \Gamma^2 \left(\frac{1}{2} (S-\Delta
   +D)\right) \Gamma (S+\Delta -1)}.
\end{align}
where the residue is computed  w.r.t. the appropriate solution of \eqref{ex-eq} for each relevant  operator. 
For instance, we can consider \(\tr(\phi_1^2)^\dagger\), which is exchanged for any even \(D\); then the perturbative expansion of \eqref{ex2} is \begin{equation}\label{struct}
C^2_{\tr\phi^2} = 2 \,+ \frac{4 i \xi}{\Gamma \left(\frac{D}{2}\right)} \left(2 \psi ^{(0)}\left(\frac{D}{4}\right)-\psi ^{(0)}\left(\frac{D}{2}\right)+\gamma_E\right)+O(\xi^2)
\end{equation}
The relations \eqref{ex-eq} and \eqref{ex2} define exact conformal data of operators propagating in the OPE channel \(x_1\to x_2\). \vspace*{5pt} 

Finally, we discuss an interesting \(D\rightarrow \infty\) limit of the theory. 
We should then rescale the coupling \(\xi=\xi_\infty \sqrt{\Gamma(D/2)}\), where \(\xi_\infty\) is fixed.  Anomalous dimension \(\gamma_\infty\) of \(\tr(\phi_1^2)\) has finite limit since for \(S=0\) in eq.\eqref{ex-eq} it is given by  \begin{align*}
-{\gamma_\infty \sin \left(\frac{\pi  \gamma_\infty}{2}\right)}=~{2 \pi }\xi_\infty^4
\end{align*} while \(\gamma_\infty\) vanishes for operators with higher spin \(S\neq 0\). As concerns the expansion \eqref{cpwe}, the number of exchanged operators becomes a countable infinity, diverging linearly in \(D\). Finally,  the OPE structure constant  \eqref{struct} for \(\tr(\phi_1^2)\)  trivially reduces to its bare value in this limit.

\section{Conclusions}
We showed that the strongly \(\gamma\) deformed \(\mathcal{N}=4\) SYM theory proposed in \cite{Gurdogan:2015csr} is just the 4-dimensional representative of a wider, \(D\)-dimensional family of theories of two complex scalar fields obtained by modifying the propagators of fields in a \(D\)-dependent way. Similarly to the \(4D\) case~\cite{Grabner:2017pgm}, {they turn out to be  conformal and integrable at any \(D\), at least in the planar limit, if we add to the action certain double-trace terms with  specific  couplings. The conformality  of our theory at finite \(N_c\) remains an open question, though it is quite plausible that the planar conformal point simply shifts to some other complex values of couplings. }  There are two such complex conjugate values of these couplings and we compute them perturbatively up to two loops. The integrability is explicit due to the domination of sufficiently large orders of perturbation theory  by the ``fishnet" Feynman diagrams.  The cylindric   fishnet graphs, related to the renormalization of ``vacuum" \(\tr(\phi_1^L)\) operators,  can be created by multiple application of a ``graph-building" operator   which appears to be an integral of motion of the integrable   conformal   \(SO(1,D+1)\)  spin chain.  We also generalize the bi-scalar model to a CFT with different propagators for the fields \(\phi_1\) and \(\phi_2\),  leading to ``non-isotropic" fishnet Feynman graphs. The underlying  graph building operator has  representations with different conformal spins in two directions on the fishnet graph.  In the  2D case the fishnet graphs are described by the same \(SL(2,\mathbb{C})\) chain as used for the dynamics of generalized Lipatov's reggeized gluons   \cite{Lipa:1993pmr} but with different value of spin, \(s=1/4\) in isotropic case, instead of the  Balitski-Kuraeev-Fadin-Lipatov reggeized gluon spin \(s=0\). This  spin chain, extensively studied in the 
literature~\cite{Korchemsky:1997fy,Korchemsky:1997fy,Derkachov:2001yn,DeVega:2001pu,Balitsky:2013npa,Balitsky:2015oux,Balitsky:2015tca},  is restored in the singular limit \(\omega\rightarrow 0\) of our bi-scalar model \eqref{bi-scalarL-alpha}.
In the spirit  of \cite{Grabner:2017pgm}, we  computed here  the exact four point correlator at any \(D\) as an expansion into conformal blocks with explicit OPE coefficients and dimensions of exchange operators in one of the channels.  In 1\(D\) case, our results are similar to the scalar version of conformal Sachdev-Ye-Kitaev fermionic theory \cite{Gross:2017aos} at \(q=4\).  For even \(D\) we found a finite, \(D\)-dependent number of  local exchange  operators at a given spin and dimension. {It would be very interesting to compute some of the discussed quantities (dimensions, structure constants) in the next \(1/N_c^2\) approximation, similarly to ~\cite{Eden2018a,Bargheer2017,Ben-Israel2018}, if the conformality of the theory holds at any \(N_c\).} The explicit form of this operators can be obtained by the  analysis of the mixing matrix for their quantum multiplets~\cite{Gromov:2017cja,Caetano:TBP}. This becomes more complicated as the dimension grows due to growing rank of the multiplets and the number of transitions, together with \(\log\)-CFT effects which arise starting from \(4D\), due to the  chirality.

Although the lagrangian \eqref{bi-scalarL-alpha} of our theory is nonlocal at general \(D\) (apart from the sequence \(D\in4\mathbb{N}\) in "isotropic" case), it  does not prevent the existence of ``normal" OPE data in this theory, which is more important for the physical interpretation of this CFT.  Moreover it would be interesting to generalize to any \(D\) the results for fishnet graphs of the type considered in \cite{Basso} and to the correlation functions for operators involving more than two scalars. Finally, an important question remains, as in \(4D\), whether these theories have any string duals at any \(D\), according to the original proposal of G~.'t~Hooft~\cite{tHooft:1973alw}. 

\begin{acknowledgments}
\section*{Acknowledgments}
\label{sec:acknowledgments}

We thank    B.~Basso, J.~Caetano,  S.~Derkachov, N.~Gromov, G.~Korchemsky, F.~Levkovich-Maslyuk   for  numerous useful discussions.  The
work of  V.K. was supported by  the
European Research Council (Programme ``Ideas'' ERC-2012-AdG 320769
AdS-CFT-solvable). The
work of E.O. is supported by the German Science Foundation (DFG)
under the Collaborative Research Center (SFB) 676 Particles, Strings and the Early
Universe and the Research Training Group 1670.

\end{acknowledgments}

\bibliographystyle{MyStyle}
\bibliography{Enrico_VK_v2}

\end{document}